\documentclass{iaus}
\usepackage{graphicx}
\usepackage{textcomp}
\title[Cosmic ray driven dynamo in barred and ringed galaxies] 
{Cosmic ray driven dynamo in barred and ringed galaxies}

\author[K. Kulpa-Dybe{\l} et al.]   
{K. Kulpa-Dybe{\l}$^1$,
K. Otmianowska-Mazur$^1$,
B. Kulesza-\.Zydzik$^1$
G.Kowal$^{1,2}$, 
D. W\'olta\'nski$^3$,
M. Hanasz$^3$  
\and K. Kowalik$^3$}

\affiliation{$^1$Astronomical Observatory, Jagiellonian
University, ul Orla 171, 30-244 Krak\'ow, Poland \break 
$^2$N\'{u}cleo de Astrofísica Teorica, Universidade Cruzeiro do Sul-Rua Galv\~{a}o Bueno 868, CEP 01506-000 S\~{a}o Paulo, Brazil \break
$^3$Centre for Astronomy, Nicholas Copernicus University, PL-87148 Piwnice/Toru\'n Poland}

\pubyear{2010}
\volume{274}  
\pagerange{119--121}
\jname{Advances in Plasma Astrophysics}
\editors{A. Bonanno, E. de Gouveia Dal Pino \& A. Kosovichev, eds.}
\begin{document}

\maketitle

\begin{abstract}
We study the global evolution of the magnetic field and interstellar medium (ISM) of the barred and ringed galaxies
 in the presence of non-axisymmetric components of the potential, i.e. the bar  and/or the oval perturbations. The magnetohydrodynamical 
dynamo is driven by cosmic rays (CR), which are  continuously supplied to the disk by supernova (SN) remnants. Additionally, 
weak, dipolar and randomly oriented magnetic field is injected to the galactic disk during SN explosions. To compare our results directly 
with the observed properties of  galaxies we construct realistic maps of high-frequency polarized radio emission. The main result is that 
CR driven dynamo can amplify weak magnetic fields up to few $\mu$G within few Gyr in barred and ringed galaxies. What is more, the modelled
 magnetic field configuration resembles maps of the polarized intensity observed in barred and ringed galaxies. 
\keywords{ galaxies: evolution, magnetic fields, ISM: cosmic rays, methods: numerical}
\end{abstract}
\firstsection 
\section{Introduction}
To explain the observational properties of the magnetic field in barred and ringed galaxies the dynamo action is necessary. 
It is thought that the CR driven dynamo can be responsible for the following effects: amplification of galactic magnetic fields up to several $\mu$G 
within a lifetime of a few Gyr; large magnetic pitch angles of about $-35$\textdegree
; symmetry (even, odd); maintenance of the created magnetic fields in a steady state; magnetic field which does not follow the gas distribution,
 i.e. magnetic fields in NGC 4736 crossing the inner gaseous ring without any change of their direction (Chy\.zy \& Buta 2008)
 or magnetic arms in NGC 1365 which are located between gaseous spiral (Beck et al. 2002).
\section{Cosmic ray driven dynamo}
CR driven dynamo is based on the following effects (details see Hanasz et al 2009 and references therein, and Hanasz et al, this volume):
The CR component described by We numerically investigated the CR driven dynamo 
model in a computational domain which covers 30~kpc $\times$ 30~kpc $\times$ 7.5~kpc of space with 300 $\times$ 300 $\times$ 75 cells of 3D
Cartesian grid, what gives 100 pc of spatial resolution in each direction. 
the diffusion-advection transport is appended to the set of resistive MHD equations. The CR energy is continuously supplied to the disk by 
SN remnants. In all models we assume that $10\%$ of $10^{51}$~erg SN kinetic energy output is converted into the CR energy, 
while the thermal energy from SN explosions is neglected. Additionally, no initial magnetic field is present but the weak 
and randomly oriented magnetic field is introduced to the disk in $10\%$  of SN explosions. Following Giacalone \& Jokipii (1999) we 
assume that the CR gas diffuses anisotropically along magnetic field lines. In order to allow the topological evolution of magnetic fields we 
apply a finite uniform magnetic resistivity of the ISM. All numerical  simulation have been performed with the 
aid of the Godunov code (Kowal et al. 2009).
\section{Barred galaxy}
\begin{figure}[t]
\begin{center}
 \includegraphics[width=0.33\columnwidth]{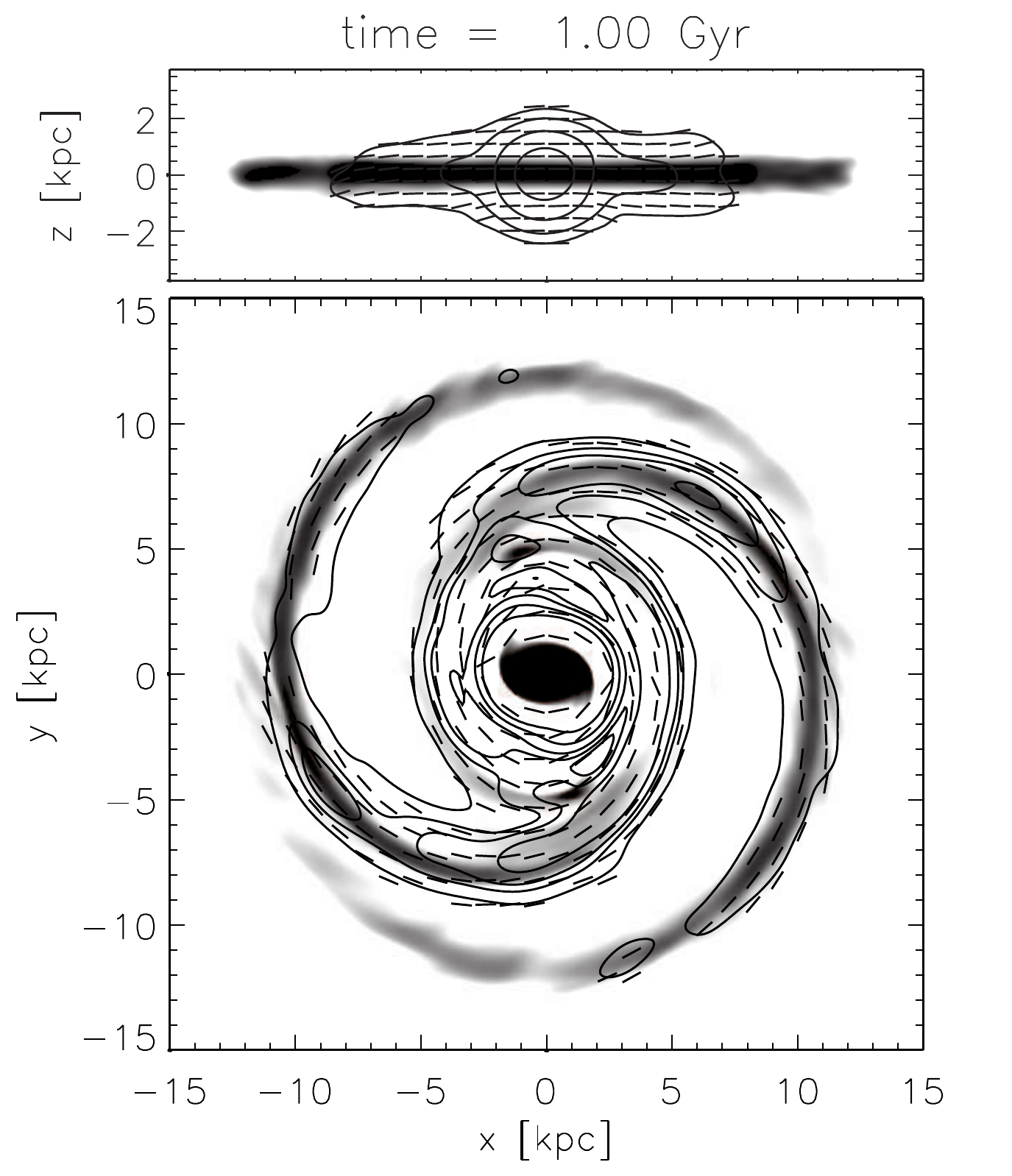}\includegraphics[width=0.33\columnwidth]{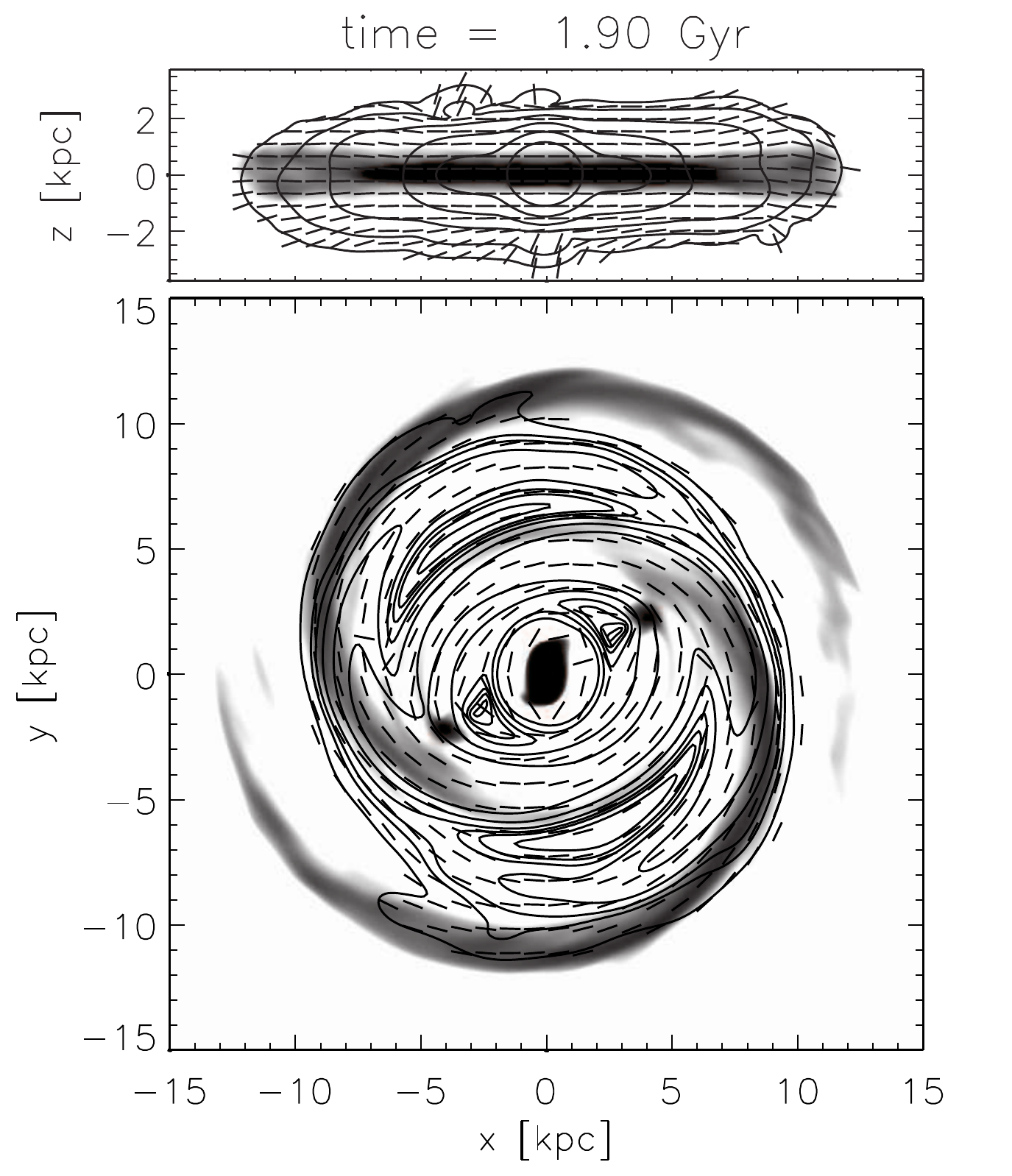}\includegraphics[width=0.33\columnwidth]{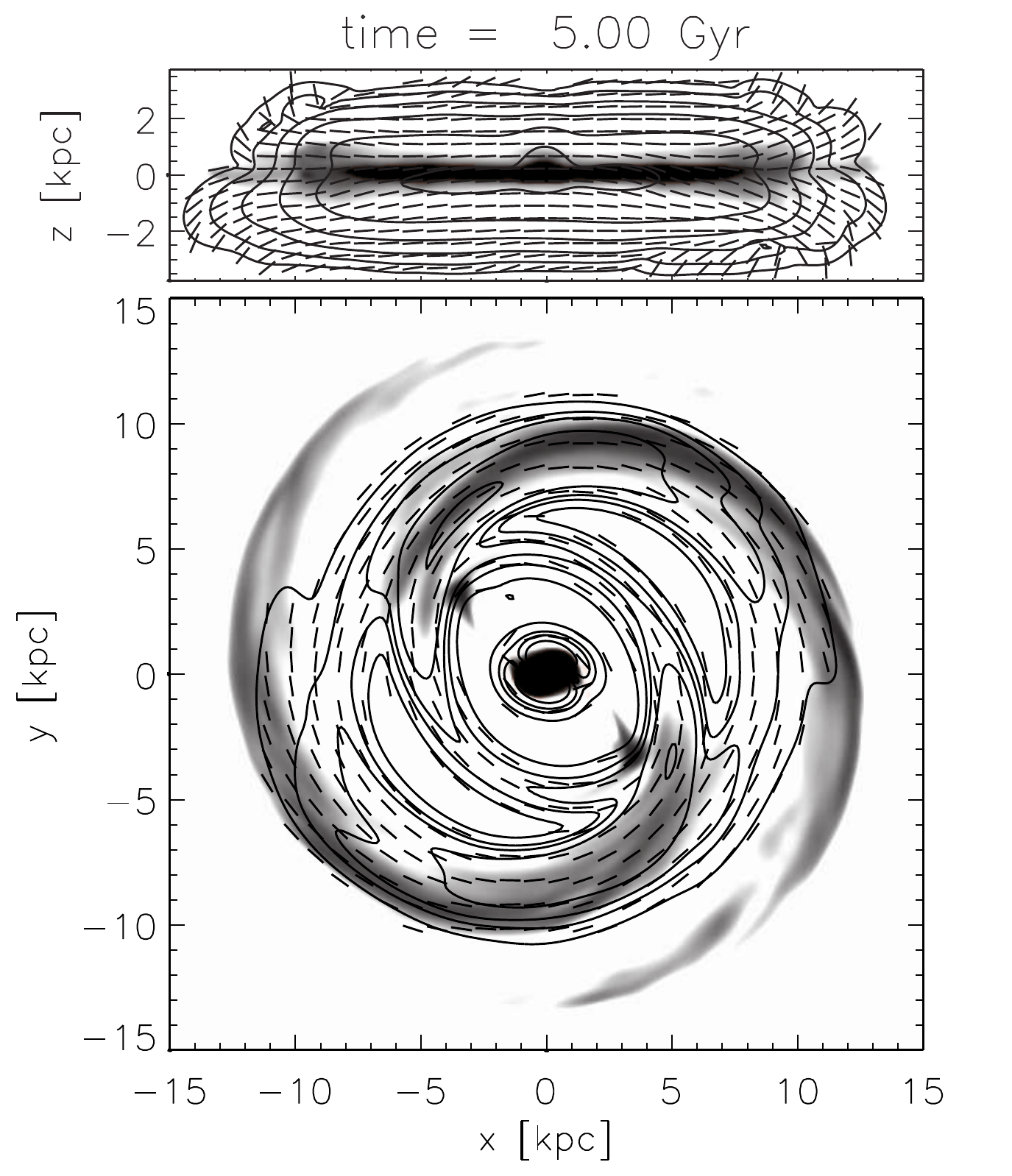}
 \caption{Face-on and edge-on polarization maps at $\lambda=6.2$~cm for selected times steps. Polarized intensity 
(contours) and polarization angles (dashes) are superimposed onto column density plots (grey-scale). All maps have 
been smoothed down to the resolution $40''$. The black color represents the regions with the highest density.}
 \label{fig:pol}
\end{center}
\end{figure}

Figure~\ref{fig:pol} shows the distributions of polarization angles (vectors) and polarized intensity (contours) 
superimposed  onto the column density maps.  On the face-on maps the magnetic field initially follows the gas distribution, 
as can easily be seen for time  $t=1.0$~Gyr, where the magnetic field strength maxima are aligned with the gaseous ones. 
However, at later time-steps, the magnetic arms begin to detach themselves from the gaseous spirals and drift into the inter-arm regions.  
In the edge-on maps the extended structures of the polarization vectors are present. This configuration of the magnetic field vectors bears 
some  resemblance to the extended magnetic halo structures of the edge-on galaxies (so called X-shaped structures).

\begin{figure}[t]
\begin{center}
 \includegraphics[width=0.33\columnwidth]{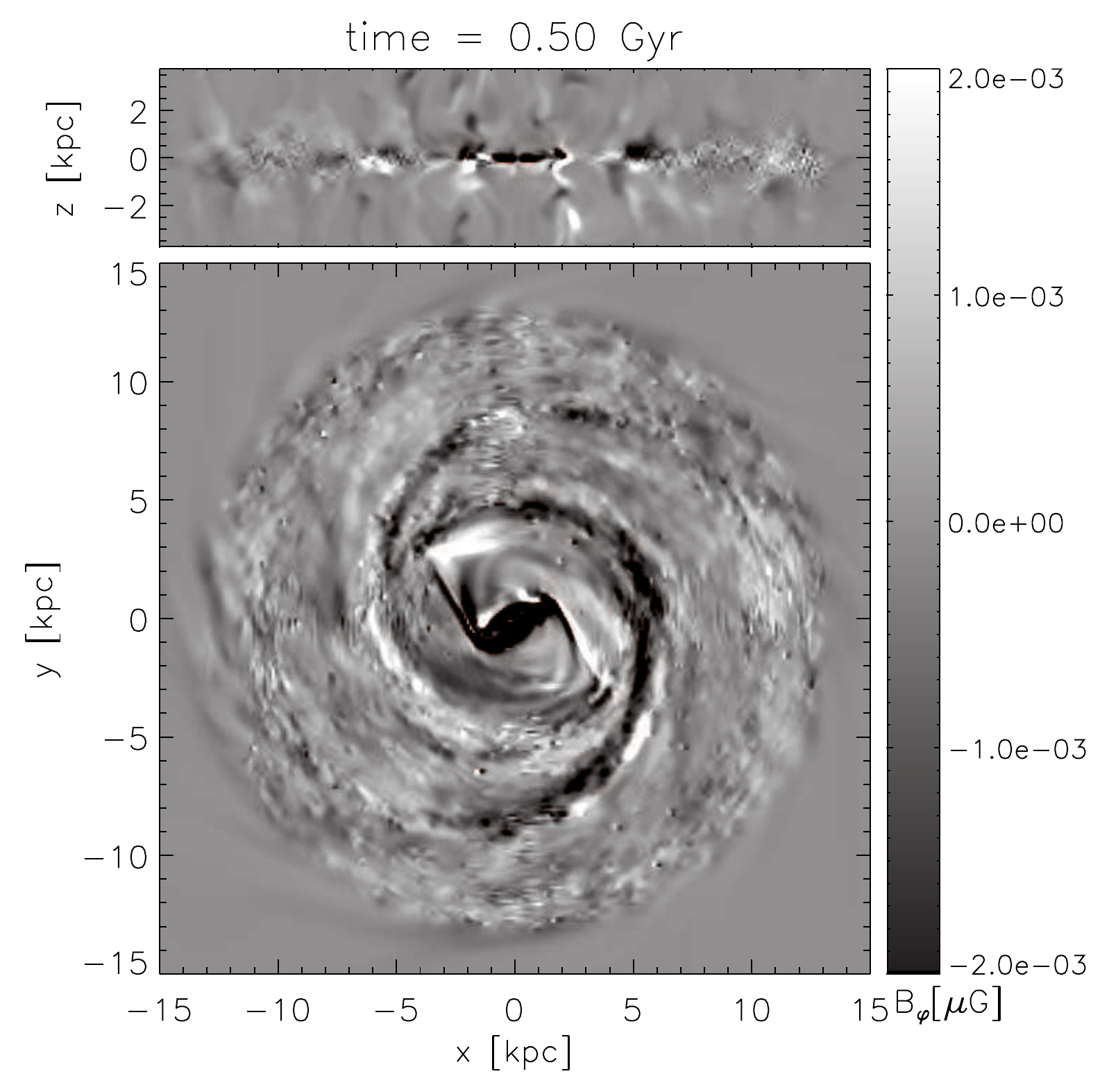}\includegraphics[width=0.33\columnwidth]{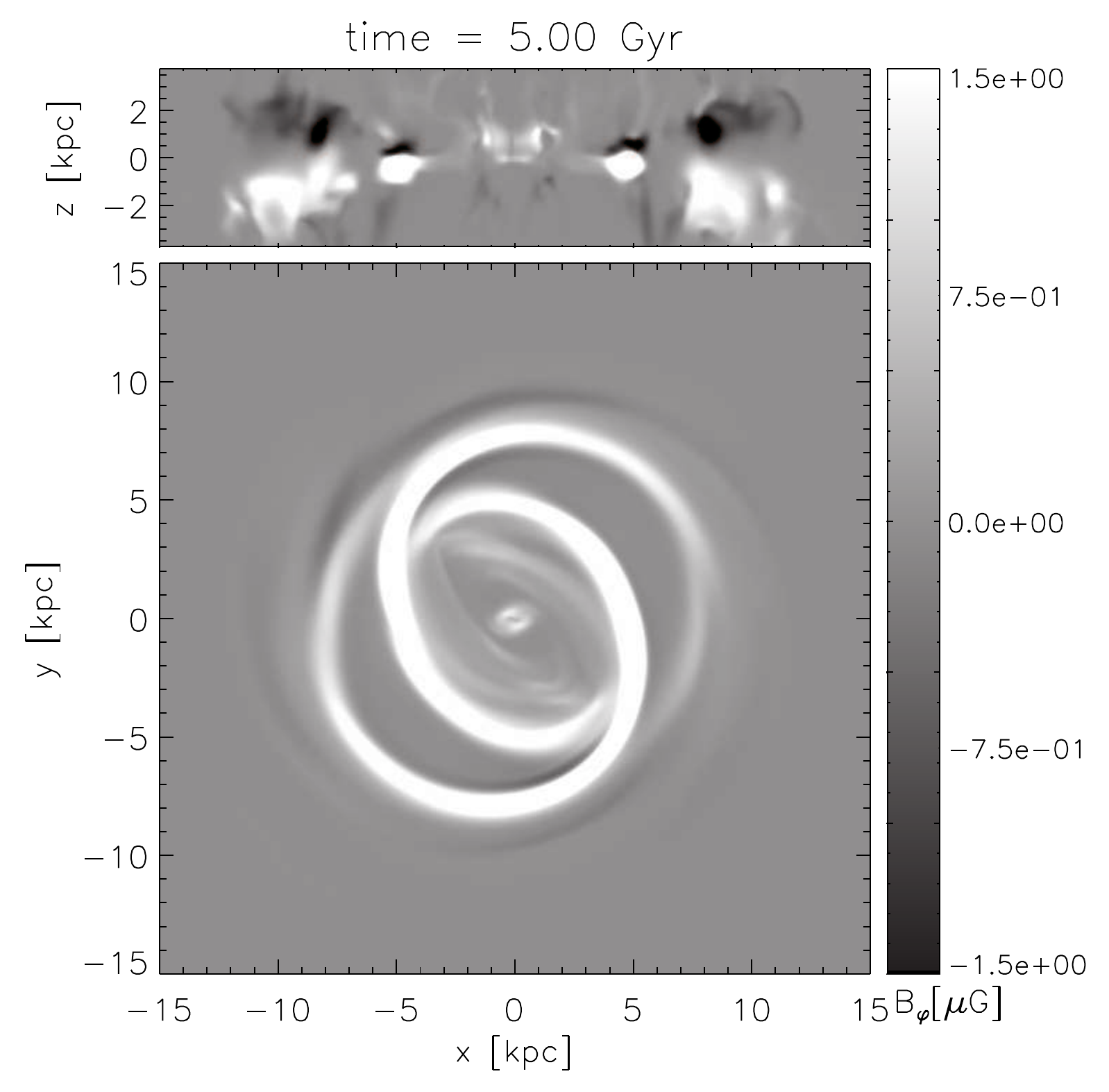}\includegraphics[width=0.33\columnwidth]{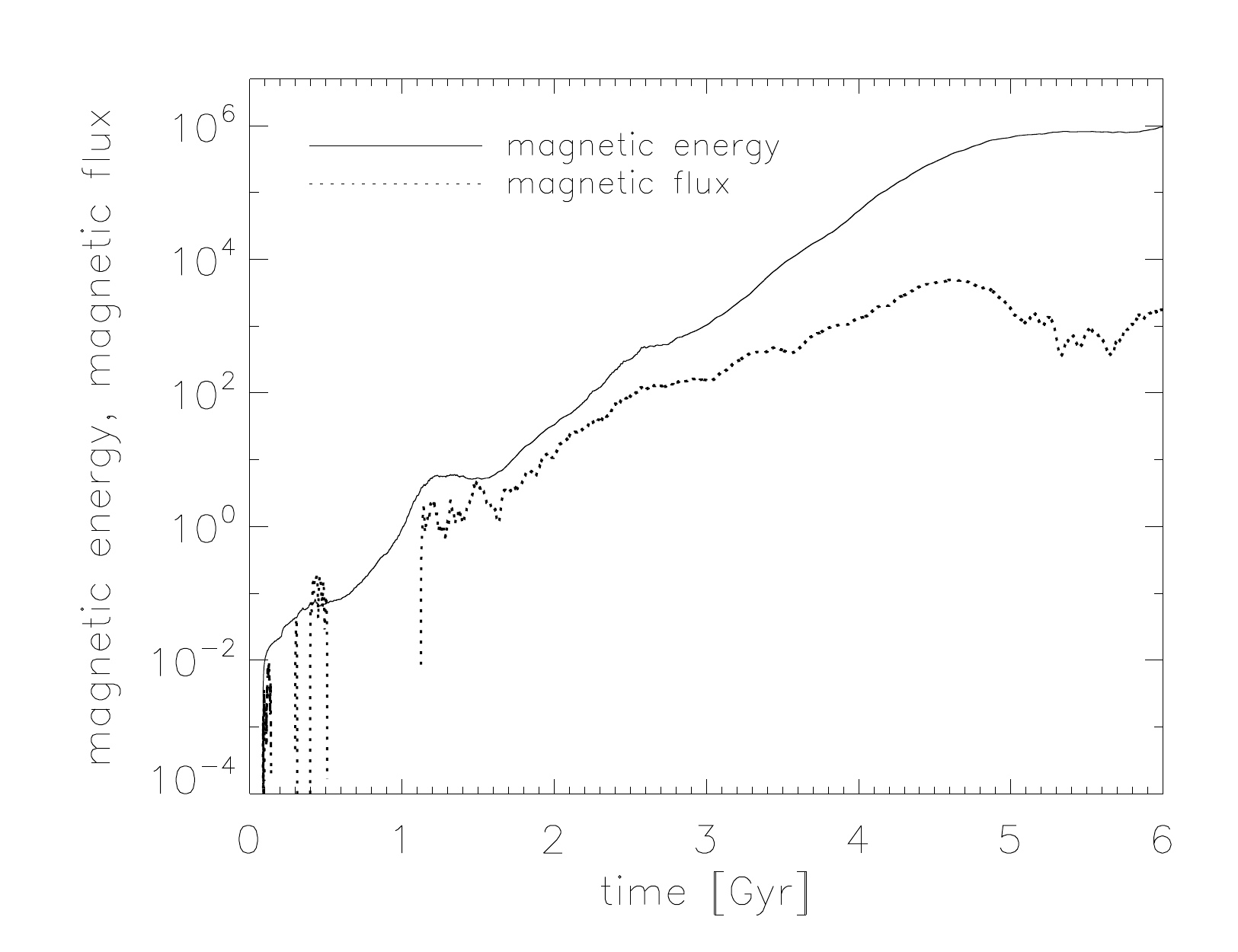} 
 \caption{\textit{Left and middle panel:} Distribution of the toroidal magnetic field for selected times steps. The white color 
represents the regions with the positive toroidal magnetic field, while black with negative. 
\textit{Right panel:} The time dependence of the total magnetic field energy $B^2$ (solid line) 
and the mean $B_\phi$ flux (dashed) calculated for the barred galaxy model.}
 \label{fig:bar}
\end{center}
\end{figure}

The face-on and edge-on distribution of the toroidal magnetic field is displayed in Figure~\ref{fig:bar} (left and middle panels). 
At $t=0.5$~Gyr the toroidal magnetic field is
 mainly disordered as it is introduced to the disk through randomly oriented SN explosions. At later time  ($t=5.1$~Gyr)  most of the toroidal 
magnetic field becomes well ordered. Moreover, we can distinguish regions with negative and positive toroidal magnetic field which form
an odd (dipole-type) configuration of the magnetic field with respect to the galactic plane.
In Figure~~\ref{fig:bar} (right panel) the exponential growth of the total magnetic energy and the total azimuthal flux due to the CR driven dynamo action
in the barred galaxy is shown.  At time $t=5$~Gyr the CR dynamo action saturates and the magnetic field reaches equipartition.
\section{Ringed galaxy}
\begin{figure}
\begin{center}
 \includegraphics[width=0.33\columnwidth]{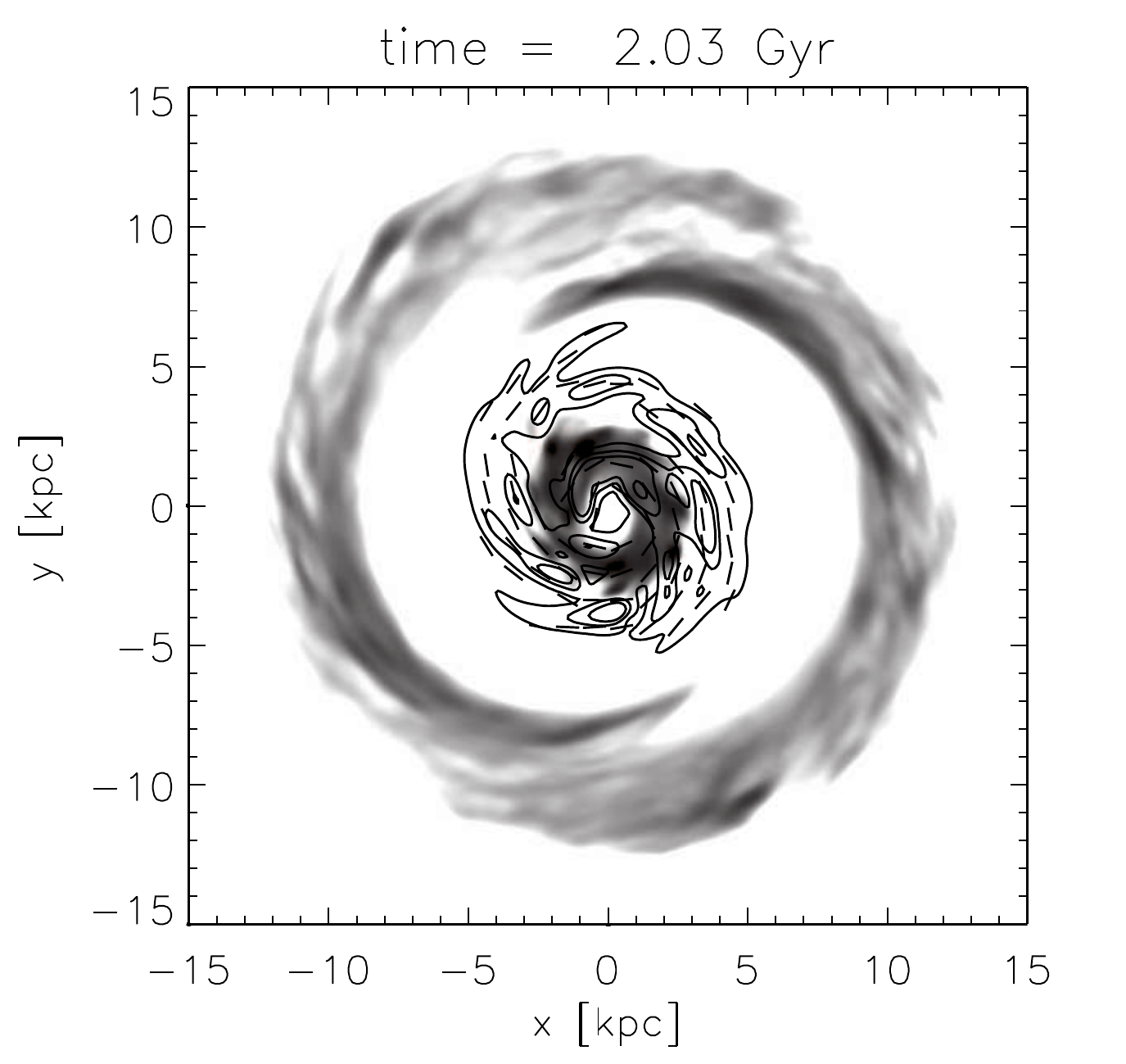}\includegraphics[width=0.33\columnwidth]{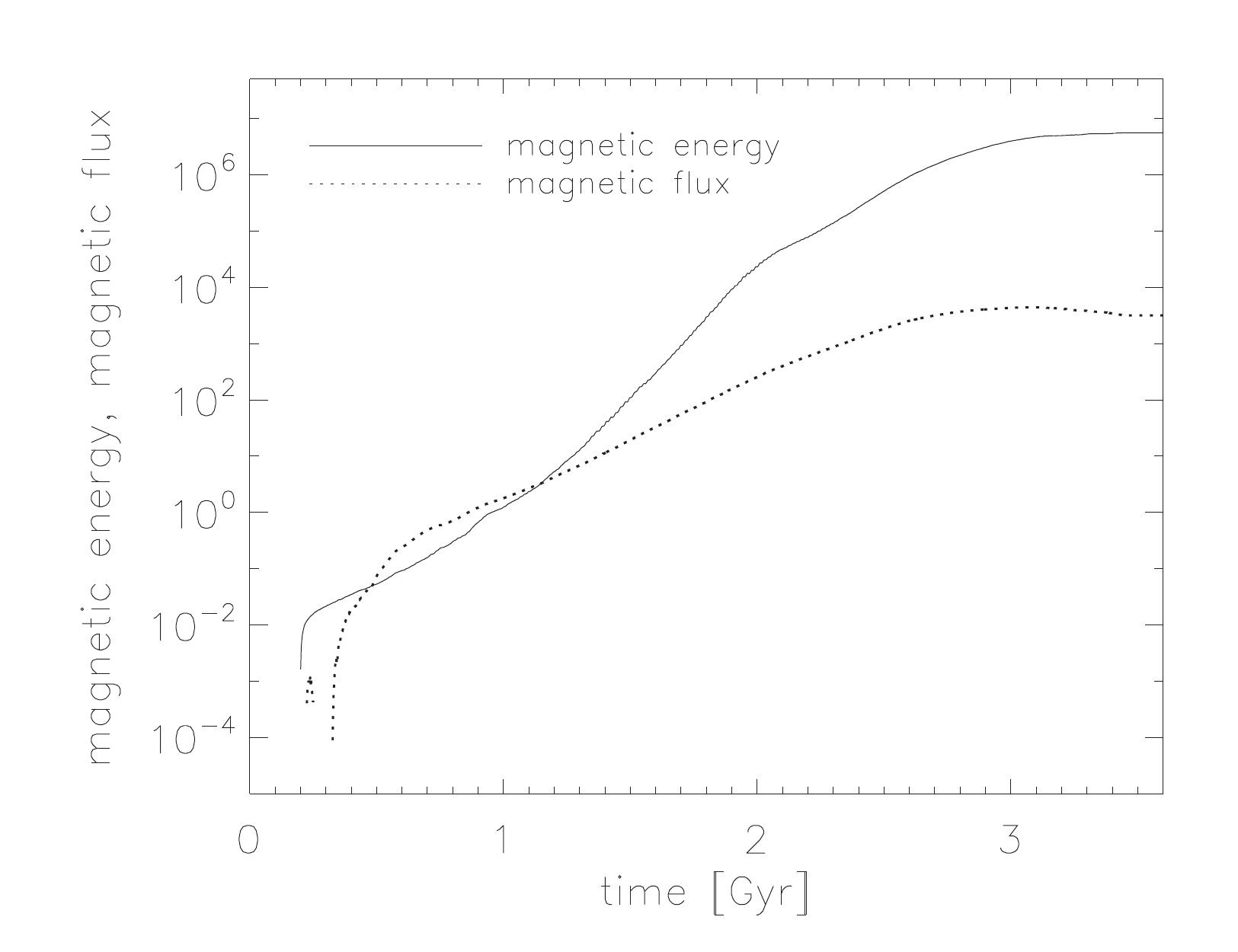} 
 \caption{\textit{Left  panel:} The face-on polarization map at $\lambda=6.2$~cm at time  $t=2.03$~Gyr 
  superimposed onto gaseous map. The map has been smoothed to the resolution $40''$. The black color represents 
the regions with the highest density, white with the smallest.
\textit{Right panel:} The time dependence of the total magnetic field energy $B^2$ (solid line) 
and the mean $B_\phi$ flux (dashed) calculated for the ringed galaxy model.}
 \label{fig:ring}
\end{center}
\end{figure}

CR driven dynamo also works in the case of the ringed galaxy NGC 4736. The exponential growth of the 
magnetic energy is even faster than in the case of the barred galaxy (Figure~\ref{fig:ring}, right panel). The obtained distribution of the gas
 density (Figure~\ref{fig:ring}, left panel)  as well as the distribution of polarization angles (vectors) and polarized intensity (contours)  bear 
some resemblance to the observation of NGC 4736. To get better results a more sophisticated numerical model of the NGC 4736 is planned.
Namely, our  model of the ringed galaxy consists of four components: the large and massive halo, the central 
bulge, the outer disc and  the oval distortion. However, following observations of the NGC~4736 this galaxy possesses one 
more component: the very small bar. Thus, to get better results we have to include the additional small bar in our simulations.
\begin{acknowledgments}
This work was supported by Polish Ministry of Science and Higher Education through grants: 92/N-ASTROSIM/2008/0 and 3033/B/H03/2008/35.
The  computations pre\-sent\-ed here have been performed on the GALERA supercomputer in 
TASK Academic Computer Centre in Gda\'nsk.
\end{acknowledgments}

\end{document}